\documentclass{mn2e}
\usepackage{times}

\input{epsf}

\def\simlt{\mathrel{\rlap{\lower 3pt\hbox{$\sim$}}
        \raise 2.0pt\hbox{$<$}}}
\def\simgt{\mathrel{\rlap{\lower 3pt\hbox{$\sim$}}
        \raise 2.0pt\hbox{$>$}}}

\begin{document}
\title[The pre-main sequence binary HK Ori]{The pre-main sequence binary HK Ori - 
Spectro-astrometry and EXPORT data}
\author[D. Baines et al.]
{\parbox[t]\textwidth{
D. Baines$^{1,2}$,   
R.D. Oudmaijer$^1$,  
A. Mora$^3$,  
C. Eiroa$^3$, 
John M. Porter$^{4}$,
B. Mer\'\i n$^5$, 
B. Montesinos$^{5,6}$, 
D. de Winter$^7$,  
A. Cameron$^{8}$,  
J.K. Davies$^9$,  
H.J. Deeg$^{10}$,  
R. Ferlet$^{11}$,  
C.A. Grady$^{12}$,  
A.W. Harris$^{13}$,  
M.G. Hoare$^1$,  
K. Horne$^{8}$,  
S.L. Lumsden$^1$,  
L.F. Miranda$^6$, 
A. Penny$^{14}$,  
A. Quirrenbach$^{15}$
}
\vspace*{6pt} \\
{\it $^1$School of Physics and Astronomy, University of Leeds, Leeds LS2 9JT, UK}\\
{\it $^{2}$U.K. Gemini Support Group, Dept. of Astrophysics, Oxford University, Keble Road, Oxford OX1 3RH,  U.K. }\\ 
{\it $^3$Departamento de F\'\i sica Te\'orica, M\'odulo C-XI, Facultad de  Ciencias, Universidad Aut\'onoma de Madrid, 28049 } \\  {\it Cantoblanco, Madrid, Spain }\\{\it $^{4}$ Astrophysics Research Institute, Liverpool John Moores University, Twelve Quays House, Egerton Wharf,}\\ {\it  Birkenhead CH41 1LD, UK }\\
{\it $^5$Laboratorio de Astrof\'{\i}sica Espacial y F\'{\i}sica Fundamental (LAEFF),  Apartado 50727, 28080 Madrid, Spain}\\
{\it $^6$Instituto de Astrof\'\i sica de Andaluc\'\i a-CSIC, Apartado 3004, 18080 Granada, Spain}\\
{\it $^7$TNO/TPD-Space Instrumentation, Stieltjesweg 1, PO Box 155, 2600 AD Delft, The Netherlands}\\
{\it $^{8}$Department of Physics and Astronomy, University of St. Andrews, North Haugh, St. Andrews KY16 9SS, Scotland, UK}\\
{\it $^9$Astronomy Technology Centre, Royal Observatory, Blackford Hill, Edinburgh, EH9 3HJ, UK}\\
{\it $^{10}$Instituto de Astrof\'\i sica de Canarias, c/Via L\'actea s/n, 38200  La Laguna, Tenerife, Spain}\\
{\it $^{11}$CNRS, Institute d'Astrophysique de Paris, 98bis Bd. Arago, 75014 Paris, France}\\
{\it $^{12}$NOAO/STIS, Goddard Space Flight Center, Code 681, NASA/GSFC, Greenbelt, MD 20771, USA}\\
{\it $^{13}$DLR Department of Planetary Exploration, Rutherfordstrasse 2, 12489 Berlin, Germany}\\
{\it $^{14}$Rutherford Appleton Laboratory, Chilton, Didcot, Oxfordshire OX11 0QX, UK}\\
{\it $^{15}$Sterrewacht Leiden, PO Box 9513, 2300 RA Leiden, The Netherlands}\\
}

%\date{Accepted Received}
%\pubyear{2004}

%\label{firstpage}

\maketitle
\vspace{5cm}

\begin{abstract} 
In this paper we present multi-epoch observations of the pre-main
sequence binary HK Ori.  These data have been drawn from the EXPORT
database and are complemented by high quality spectro-astrometric data
of the system.  The spectroscopic data appear to be very well
represented by a combination of an A dwarf star spectrum superposed on
a (sub-)giant G-type spectrum.  The radial velocity of the system is
consistent with previous determinations, and does not reveal binary
motion, as expected for a wide binary.  The spectral, photometric and
polarimetric properties and variability of the system indicate that
the active object in the system is a T Tauri star with UX Ori
characteristics.  The spectro-astrometry of HK Ori is sensitive down
to milli-arcsecond scales and confirms the speckle interferometric results
from Leinert et al. The spectro-astrometry allows with fair certainty
the identification of the active star within the binary, which we
suggest to be a G-type T Tauri star based on its spectral characteristics.
\end{abstract}


\begin{table*}
\caption{{Log of the observations. 
Optical photometry and polarimetry were obtained at the NOT, and the
near-infrared photometry at the CST. Spectroscopy was obtained at the
INT, WHT and the AAT. The symbol `--' indicates non photometric
nights, while fields are left blank when the object was not
observed. Typical errors on {\it V} and {\it K} are about 0.05 mag,
and 0.1\% in polarization. Error bars on the equivalent widths are of
order 5-10\%. See text for details.}
\label{log}}
\begin{tabular}{rcccccccc}
\hline
\hline
UT Date & Julian Date & Telescope & Exp. time & {\it V}&{\it P$_V$}& {\it K} & H$\alpha$ \\
        & 2450000+ &      &   (min)   &  (mag) &  (\%)       & (mag)  & EW ($\rm \AA$) \\
\hline
\hline
24-10-1998 & 1110.72 &     &     &  11.59 & 0.96 & 7.23  & \\
25-10-1998 & 1111.68 & INT & 20  &  11.69 & 1.22 & 7.23  & --59.2 \\
26-10-1998 & 1112.74 & INT & 10  &  11.71 & 1.26 & 7.25  & --57.8 \\
27-10-1998 & 1113.66 & INT & 10  &  11.68 & 1.45 & 7.21  & --63.6 \\
28-10-1998 & 1114.71 & INT & 10  &  11.79 & 1.74 &       & --57.8 \\
29-10-1998 & 1115.68 & INT & 10  &        &      &       & --63.7 \\

29-01-1999 & 1208.49 & INT & 15  &   --   & 0.86 & 7.18  & --54.0 \\
31-01-1999 & 1209.51 & INT & 15 &  11.44 & 0.95 & 7.26  & --55.1 \\
01-02-1999 & 1210.50 & INT & 15  &        &      & --    & --58.2 \\
\hline
30-01-1999 & 1209.46 & WHT & 45  &        &      &       &    \\
\hline
29-01-2002 & 2303.99 & AAT & 8 $\times$ 5m &      &      &       &   --44    \\
\hline
\hline
\end{tabular}
\ , \\
\end{table*}

\section{Introduction}

Multiplicity amongst pre-main sequence stars is at least as common as
that in main sequence stars, or perhaps even higher as found for
Herbig Ae/Be stars (Bouvier \& Corporon 2001, for a review on binary
formation see Tohline 2002). The study of young binaries should
therefore provide further clues to the star formation process.  In
this paper we present our results on HK Ori, an intriguing young
binary system. Initially classified as having a peculiar A4 emission
type spectrum by Joy (1949), it was later proposed to be a Herbig
Ae/Be star (Herbig, 1960).  Other spectral type classifications have
been published, for example, Mora et al. (2001, and references
therein) classify HK Ori as a G1Ve type star based on data in the
wavelength region 5700 - 6100 $\rm \AA$. That different spectral types
were derived when different wavelength ranges were used, was initially
suggested to be due to a spectroscopic binary system (Finkenzeller \&
Mundt, 1984; see also Corporon \& Lagrange 1999, and Mora et al.,
2001). Spectral types based on blue spectra tend to result in earlier
classifications, while neutral metallic lines, indicative of a cool
star, begin to dominate in the red. Recently, it was proven that HK
Ori is indeed a binary system.  Using speckle interferometry in the
near infrared, Leinert, Richichi \& Haas (1997) found a binary
separation of 0.34$\pm$0.02 arcsec, corresponding to 156 AU at an
assumed distance of 460 pc (Leinert et al.).  There is a large colour
difference: both objects are of almost equal brightness at 1 $\mu$m,
rising to a difference of 2.3 mag in the $K$ band. Understanding the
nature of the HK Ori system is dependent on the trivial, but
unanswered, question on whether the A-type or the G-type star
dominates at {\it K} wavelengths.  Conventional wisdom has it that if
only one pre-main sequence star emits excess radiation in the
near-infrared, this has to be the late type star.  This is because the
contraction timescale for lower mass stars is longer, they will
therefore arrive later on the main sequence than more massive objects.
Leinert et al.  (1997) suggested two possible models for the binary HK
Ori system.  The first model assumes that the object dominating the
infrared radiation is also the brighter star in the optical, while
conversely, for the second case they assume that one star is
dominating the infrared and the other star the optical
radiation. Neither of the two possibilities could be excluded on the
basis of their data.

HK Ori was one of the targets in the EXPORT (Exo-Planetary
Observational Research Team, Eiroa et al. 2000) collaboration.  This
consortium obtained the `1998 La Palma International Time', which
amounts to 5\% of observing time on all La Palma and Tenerife
telescopes for one year. In this paper we describe the findings on HK
Ori. In section 2 we present the EXPORT data and the new
spectro-astrometric observations. The method used to extract
information from such data is outlined in section 2.2.  The data will
be used to confirm the presence of a binary and to discriminate
between Leinert et al.'s (1997) two scenarios.  Section 3 contains the
results deduced from the EXPORT data. In section 4 the results on
spectro-astrometry are presented. Section 5 contains the discussion
and some final remarks.

\section{Observations} 

During the 1998 La Palma International Time campaign, the EXPORT
collaboration obtained multi-wavelength observations of young stars on
16 observing nights. A major aim of the EXPORT project is to observe a
large sample of Herbig Ae/Be stars to learn about their temporal
variability.  During four runs in May, July, October 1998 and January
1999, 72 target stars were observed spectroscopically at the 2.5m
Isaac Newton Telescope (INT) and the 4.2m William Herschel Telescope
(WHT) and photo-polarimetrically at the 2.5m Nordic Optical Telescope
(NOT) and the 1.5m Carlos S\'anchez Telescope (CST). From these data
we have available for HK Ori eight medium resolution red spectra taken
on timescales of days and months, one high resolution blue spectrum
and multi-epoch photo-polarimetric data coincident with the
observations of the red spectra. The INT, WHT and NOT are located at
the Observatorio del Roque de los Muchachos, La Palma, and the CST is
at the Observatorio del Teide, Tenerife.

We also obtained spectro-astrometric data on the 3.9m Anglo-Australian
Telescope (AAT) at Siding Spring, Australia.  With these data we are
able to probe structures at scales of milli-arcseconds (see below and
Bailey 1998).

\subsection{EXPORT data}

The intermediate resolution spectra were taken with the INT equipped
with the Intermediate Dispersion Spectrograph (IDS). An EEV CCD was
used during the observations runs of HK Ori. The wavelength coverage
was 5712 to 6812 $\rm \AA$, with $\sim$0.475 $\rm \AA$ per pixel and a
resolving power of $\sim$6600 at 6300 $\rm \AA$. The slit width was
always 1.0 arcsecond projected on the sky.

The high resolution spectrum was taken on 30 January 1999 using the
Utrecht Echelle Spectrograph (UES) on the WHT. The instrumental set-up
resulted in a complete spectrum from $\sim$3700 to $\sim$6100 $\rm \AA$,
with the intermediate and high resolution spectra overlapping between
5700 and 6100 $\rm \AA$, where the He {\sc i} 5876 and Na {\sc i} D
lines are located.  The spectra were dispersed into 59 echelle orders
with a resolving power of 49000 and a slit width of 1.15 arcsec
projected on the sky.  We refer to Mora et al. (2001) for further
details on both the INT and WHT observing procedures and data
reduction.

{\it UBVRI} photometry and polarimetry were obtained simultaneously
using the Turpol {\it UBVRI} polarimeter/photometer mounted on the
NOT.  Six photometric and seven polarimetric observations were taken
in October 1998 and January 1999. Only the night of 29 January was not
photometric.  More details about these observations are provided in
Oudmaijer et al. (2001).  Seven near-infrared {\it JHK} photometric
data points were obtained, almost simultaneously with the INT spectra,
on the CST. For more details on the near-infrared data, we refer to
Eiroa et al. (2001).

Table \ref{log} shows the log of the observations. Columns 1 and 2
give the dates and JD's of the spectroscopic observations, columns 3 
and 4 give the telescopes where these were obtained and the exposure
times, columns 5, 6 and 7 give the values of {\it V}, polarization and
{\it K}, and column 8 the equivalent width of H$\alpha$. 
 
\subsection{Spectro-astrometry}

As the technique is not widely used, we will briefly outline the
method, mainly following Bailey (1998). The basic idea of
spectro-astrometry is fairly simple, it measures the relative spatial
position of spectral features from a longslit spectrum. For example,
if one observes a binary system with a large separation in a
2-dimensional longslit spectrum, we would detect two easily
distinguishable spectra. In the case of a close, unresolved, binary,
we would detect only one spectrum.  However, consider now the
situation that one of the two stars dominates at a given wavelength,
for example at H$\alpha$.  The continuum traces the combined emission
from both stars, but across the emission line the spatial position of
the unresolved spectrum shifts in the direction of the H$\alpha$
emitting star. In this manner, the presence of a binary can be
revealed.  For a well sampled, high signal-to-noise spectrum, we can
measure the centroid of the spatial profile to a fraction of a pixel,
often down to orders of milli-arcseconds.  In order to detect any
binary, it is necessary to obtain observations at (at least) 2
different slit angles. If no prior knowledge is present, the obvious
choices are the North-South and East-West directions. Usually, to
account for systematic effects, observations are done at 4 position
angles; 0$^{\rm o}$, 90$^{\rm o}$, 180$^{\rm o}$, and 270$^{\rm o}$
equivalent to N-S, E-W, S-N and W-E.

Bailey (1998) already showed the power of the method when he
discovered sub-arcsecond binaries. Takami, Bailey \& Chrysostomou
(2003 and references therein), studied a large sample of T Tauri stars
and described circumstellar material at AU scales (see also Whelan,
Ray \& Davis 2004 for a different application in the Pa$\beta$
line). Independently, we had started a spectro-astrometric survey of
31 Herbig Ae/Be stars (described in Baines 2004, see also Porter,
Oudmaijer \& Baines 2004) in which HK Ori was included.

Observations of HK Ori were carried out on 29 January 2002 at the AAT
using the RGO spectrograph with its 82 cm camera and a MITLL 2048
$\times$ 4096 CCD. A 1200 line mm$^{-1}$ grating was used to give a
wavelength coverage of about 6280 to 6770 {\rm \AA} for all
observations.  A 1 arcsec slit width resulted in a resolving power of
about 7500 (40 km s$^{-1}$) at H$\alpha$.  The instrumental set-up
gives a pixel size of 0.15{\rm \AA} in the dispersion direction and a
spatial pixel size of 0.15 arcsecond, amply sampling the spectral
resolution and the seeing ($\sim$2 arcsec during the observations of
HK Ori) respectively.  Exposures of a CuAr lamp were taken for
wavelength calibration.  As mentioned, the spectra were obtained at
four slit position angles to correct for any instrumental effects such
as curvature or optical distortion introduced by the spectrograph,
misalignment of the spectrum with the CCD columns, or any departure of
the CCD pixels from a regular grid.

The data were reduced using the IRAF package. High signal-to-noise
ratio flat fields were made by combining many exposures with the
spectrograph illuminated by a tungsten lamp. After subtracting the
bias level and dividing by a normalised flat field, the 2-dimensional
spectrum was fitted by Gaussian profiles in the spatial direction at
each wavelength step. This resulted in a so-called position spectrum,
which gives the centre of the emission as a function of wavelength.
Four position spectra were obtained at the four different position
angles. Any instrumental effects are then largely eliminated by
averaging those with opposite position angles (0$^{\rm o}$-180$^{\rm
o}$, or 90$^{\rm o}$-270$^{\rm o}$).  The results are two position
spectra, one in the North-South (NS) direction and the other in the
East-West (EW) direction. Each position spectrum has an arbitrary
zero-point, which is adjusted to correspond to the continuum position.
The rms accuracy in the positioning of the centroid was found to be 10
mas in a seeing of 2 arcsec. HK Ori was one of the fainter targets in
our sample, rms variations of order 2 mas were routinely observed for
brighter targets (Baines 2004, Baines et al., in preparation).

\begin{figure}
\mbox{\epsfxsize=0.48\textwidth\epsfbox[40 160 570 430]
{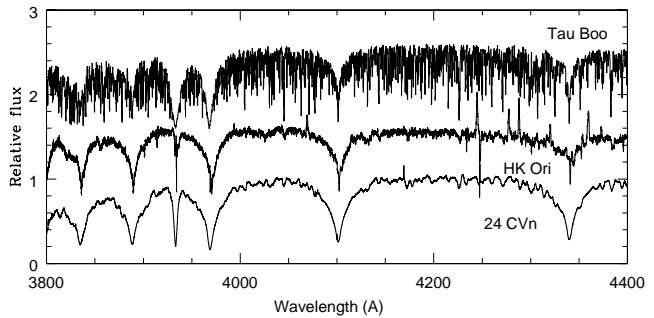}}
 \caption{Part of the UES spectra taken during the EXPORT observations. HK Ori
(middle) is plotted next to $\tau$ Boo (above, a G star) and 24 CVn
(below, A-type).  The blue spectrum of HK Ori clearly resembles that
of an A star.
\label{ueswht}
}
\end{figure}

\section{Results - EXPORT data}

\subsection{Photometry and polarimetry}

Oudmaijer et al. (2001) and Eiroa et al. (2001, 2002) describe the
optical photo-polarimetry and near-infrared EXPORT data respectively.
Here we briefly mention the observed properties of HK Ori. During
October 1998, the {\it V} magnitude faded roughly by 0.2 mags, whilst
keeping the same {\it B -- V} colour to within the errors.  However,
the observed polarization increased significantly, from $\sim$~0.8\%
to 1.8\% in the {\it V} band. The near-infrared data did not
change. This behaviour is consistent with that expected for the
so-called UX Ori stars (see e.g. Grinin, Kolotilov \& Rostopchina
1995), a dust cloud orbiting the star enters the line of sight
increasing the extinction, while increasing the polarization fraction
of the observed light at the same time. This mechanism also results in
an associated reddening, but given the small amount of dimming (0.2
mag in {\it V}), changes in {\it B--V} may not be observable within
our error bars.

HK Ori was brighter and bluer in January 1999. A comparison with the
multi-epoch data of Herbst \& Shevchenko (1999) indicates that HK Ori
was then in its brightest state.  The polarization was low, which in
principle would be consistent with the above UXOR explanation.
However, we note that in the case of UXORs, day-to-day variations in
photo-polarimetry and month-to-month variations can not easily be
compared since different dusty clouds orbiting the star may be
responsible for the polarization (e.g. Oudmaijer et al. 2001).  In
addition, as will be discussed in the next subsection, there is
evidence that the spectrum of HK Ori was veiled by blue excess
emission during the 1999 observations, hampering a proper comparison
of the polarization states of the star.

\subsection{The spectrum}

HK Ori has a complex spectrum over the wavelength interval $\sim$3700
to 6800 $\rm \AA$. The spectrum is dominated by forbidden lines and,
to a smaller degree, permitted emission lines of neutral
metals. Strong and broad double peaked H$\alpha$ and H$\beta$ emission
lines are present, with the peak separation in H$\beta$ larger than in
H$\alpha$. The He {\sc i} 5876 $\rm \AA$ is also broad and displays an
inverse P Cygni profile.  An illustration of the early spectral type
is provided in Fig.~\ref{ueswht}, where part of the WHT/UES spectrum
of HK Ori is compared with those of 24 CVn (an A star) and $\tau$ Boo
(G-type). It is clear that the blue spectrum reveals an A-type object.
Neutral metallic absorption lines (Ca {\sc i} and Fe {\sc i}) begin to
dominate the spectrum from $\sim$6000 $\rm \AA$ onwards.  The strong
Li {\sc i} 6708 \AA{} absorption indicates that a cool and young
object is also present.

\begin{figure}
\mbox{\epsfxsize=0.48\textwidth\epsfbox{
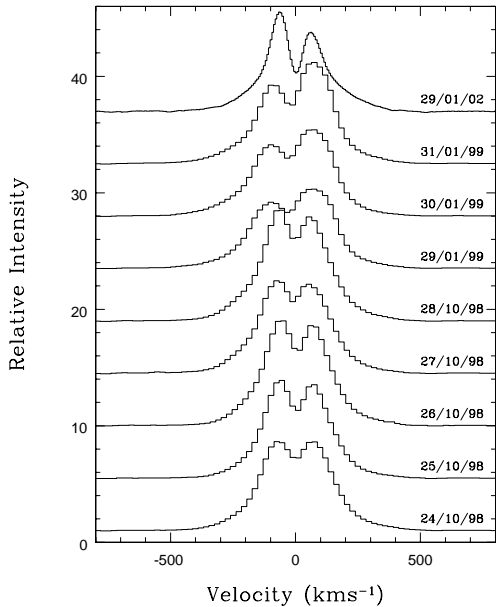}}
 \caption{H$\alpha$ line profiles of HK Ori, taken during 24-28
 October 1998, 29-31 January 1999 and 29 January 2002. The velocity is 
with respect to the systemic velocity (21 km s$^{-1}$ LSR).
\label{halpha}
}
\end{figure}

Figure~\ref{halpha} shows a plot of the medium resolution INT spectra,
centred on H$\alpha$, for all nights in October 1998 and January 1999.
For completeness, the 2002 AAT spectrum is shown as well. Both night
to night, and month to month variations appear in the profiles.  The
H$\alpha$ profiles in January are similar to the profiles found by
Grady et al. (1996), Reipurth Pedrosa \& Lago (1996), and Finkenzeller
\& Mundt (1984).  In January 1999 the profiles appear broader and
shallower than the October profiles, with a maximum velocity of
400-500 km s$^{-1}$ in the red wings, with respect to the systemic
velocity, and a V/R ratio decreasing daily to a minimum of 0.8. The
central absorption dip seems to be redshifted by about 20 kms$^{-1}$
with respect to the systemic velocity (21 km s$^{-1}$ LSR, see section
3.3). The equivalent widths are comparable in October and January at
$\sim -60$ \AA{} and $\sim -56$ \AA{} respectively, this has decreased
to $-44$ \AA{} in 2002.

\begin{figure}
\mbox{\epsfxsize=0.48\textwidth\epsfbox{
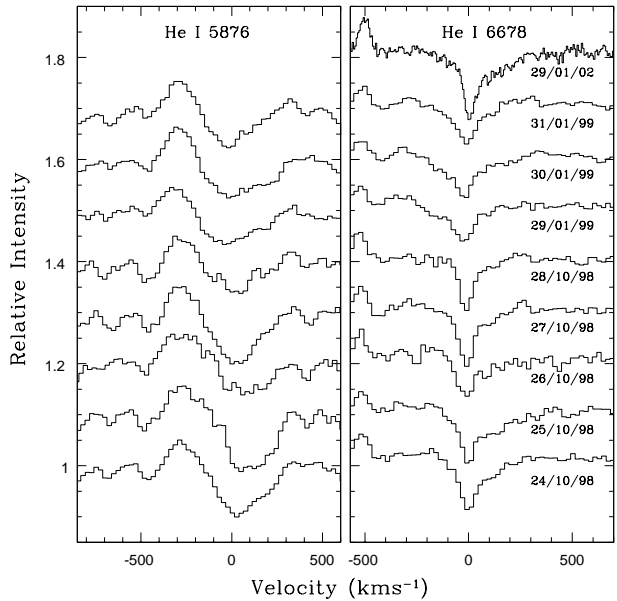}}
 \caption{The He {\sc i} 5876 and 6678 $\rm \AA$ line profiles of HK
 Ori from Both EXPORT runs. The He {\sc
 i} 6678 $\rm \AA$ line appears to be blended with the narrow Fe {\sc
 i} 6670 absorption line. The velocity scale is with respect to the 
systemic velocity.
\label{helium}}
\end{figure}

Both He {\sc i} 5876 and 6678 $\rm \AA$ are present in the spectra on
all nights. He {\sc i} 5876 displays a broad variable inverse P Cygni
(IPC) profile throughout October and January, with a maximum blueshift
velocity of -360 km s$^{-1}$ on 29 January 1999, and a maximum
redshift velocity of 270 km s$^{-1}$ on 25 October 1998. The strength
of He {\sc i} 5876 gradually decreases through October.  The He {\sc
i} 6678 $\rm \AA$ line is strong when compared to the He {\sc i} 5876
$\rm \AA$ line.  The velocity of the He {\sc i} 6678 $\rm \AA$ line
remains constant at 16 km s$^{-1}$.  Corporon \&
Lagrange (1999) suggest that the broad He {\sc i} 6678 $\rm \AA$ line
is blended with the Fe {\sc i} 6680 $\rm \AA$ absorption line from the
cooler companion star. This appears to be confirmed in our spectra
since, as can be seen in Fig.~\ref{helium}, the He {\sc i} line
profiles do not at all resemble one another. He {\sc i} 4026 and 4471
$\rm \AA$ absorption lines are also weakly present in the high
resolution spectrum. The lower excitation Na {\sc i} D absorption are
also present in the spectrum.  Their redshifted absorption wings had
previously been seen by Grady et al. (1996) and Finkenzeller \& Mundt
(1984).

In the October observations, the absorption lines across the entire
spectrum are much stronger than in January.  In the spectra that were
taken within days in October 1998 and January 1999, the absorption
lines do not vary significantly in shape or strength. However, they
seem to be variable on longer timescales.  Li {\sc i} 6708 and many
other lines have smaller equivalent widths by about 20-30\% in the
January 1999 data. As all lines still seem to be present, a change in
spectral type of the star is not very likely.  Instead, it can be
explained by introducing an extra, almost featureless continuum
veiling of 0.25.  The star is brighter and bluer in January,
consistent with the idea that a featureless, hot, black body continuum
is overlaid over the stellar spectrum.

\begin{figure}
\mbox{\epsfxsize=0.48\textwidth\epsfbox{
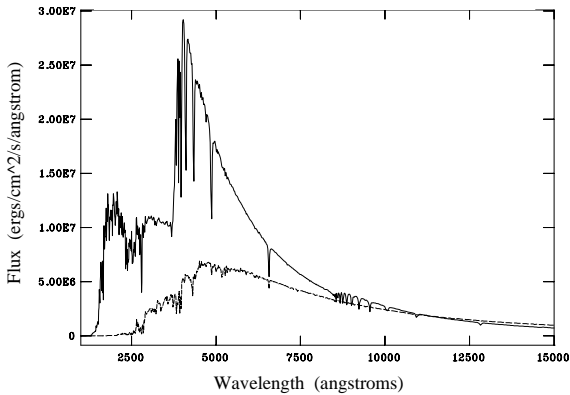}} 
\caption{Kurucz models of an A4 star (solid line) and
 a G1 star (dotted line). The flux of the G1 star is multiplied by
 2.25 so the SEDs of both stars cross at approximately 1 $\mu$m and
 1.25 $\mu$m, as observed by Leinert et al. (1997).
\label{kurucz}}
\end{figure}

\subsection{Radial velocity}

From the Fe {\sc i}{\sc i} and [O {\sc i}] emission lines
% and the strong non-variable Na {\sc i} D absorption, 
we measure a local
standard of rest (LSR) velocity of 22 $\pm$ 3 km s$^{-1}$. This value
is indicative of the radial velocity of the system since these lines
originate in the circumstellar material. As a cross check, and due
to the large number of stellar spectra obtained by EXPORT, the
standard stars of 51 Peg and $\tau$ Boo (both having well defined
radial velocities and spectral types G2.5 and G4 respectively) were
cross correlated against the spectrum of HK Ori. A radial velocity of
21 $\pm$ 3 km s$^{-1}$ was obtained and, since this matches that of
the above spectroscopic lines, will be used for the radial velocity of
the system.

A literature search returned few determinations of the radial
velocity, and does not indicate any variability.  A radial velocity of
26 $\pm$ 18 km s$^{-1}$ was found by Reipurth et al. (1996) based on
measurements of the velocity of the associated molecular
cloud. Finkenzeller \& Jankovics (1984) found a radial velocity of 20
$\pm$5 km s$^{-1}$ from the interstellar Na {\sc i} D absorption
lines by assuming these lines are formed close to the star.  These
values compare well with our own radial velocity of 21 $\pm$ 3 km
s$^{-1}$, suggesting that the system does not have a variable radial
velocity on timescales less than $\sim$20 years.

\begin{figure*}
\mbox{\epsfxsize=0.8\textwidth\epsfbox{
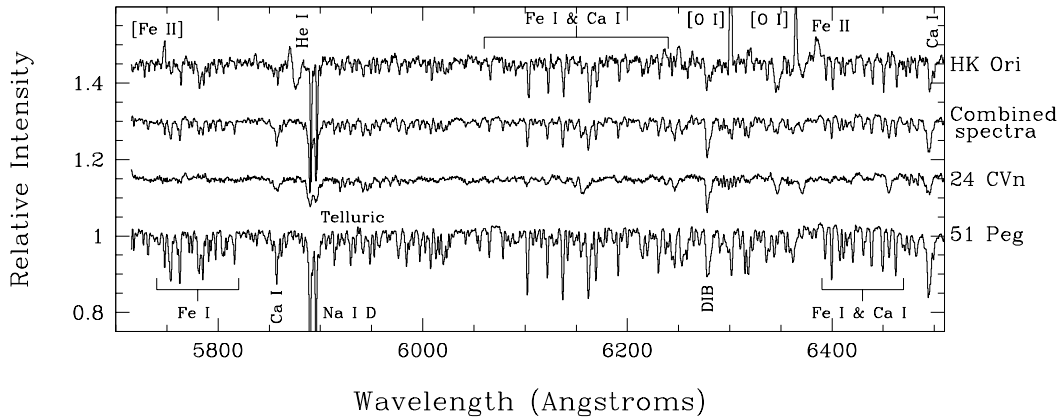}}
 \caption{The spectrum of HK Ori, 24 CVn and 51 Peg (all taken on
 29 January 1999) compared with a combined spectrum of 24 CVn and 51
 Peg in the relevant ratios found from the Kurucz model SEDs.
\label{combo}}
\end{figure*}

Since HK Ori has been found to be a binary system, it would be
interesting to know the expected radial velocity variations due to
binary motion. Using the separation of the two stars (156 AU, Leinert
et al., 1997) and assuming that the system consists of a 1 M$_\odot$ G
star and a 2 M$_\odot$ A star (as shown later), a maximum value for
the radial velocities of both stars can be calculated.  Assuming the
system to be edge-on, we obtain upper limits to the changes in the
radial velocities of 2.8 km s$^{-1}$ and 1.4 km s$^{-1}$ for the G and
A star respectively, with a binary period of $\sim$1130 years.  The
long period, combined with the small expected velocity variations, are
well within the above-mentioned observational limits. The difference
in velocities of the two stars is small as well.  This is consistent
with the fact that no difference can be found between the G star and A
star line velocities in the HK Ori spectrum.

\subsection{On the spectral type(s) of HK Ori}

For the interpretation of the spectra it is useful to know the
flux contribution from each star in the binary system HK Ori at any
wavelength within the observed region. We assume that the system is
composed of an A4 star and a G1 star since both of these spectral
types are quoted in the literature and are observed in different
spectral ranges. Due to its larger temperature, the A star is much
brighter than the G star, particularly in the blue, and for any
A+G main sequence spectroscopic binary system, the A star would easily
dominate the overall spectrum.

We produced Kurucz (1993) model atmospheres of an $\sim$A4 V star
(T$_{\rm eff}$ = 8500, log g = 4.0) plotted with a model $\sim$G1 V
star (T$_{\rm eff}$ = 5750, log g = 4.5) from 1000 to 15,000 $\rm \AA$
in Fig.~\ref{kurucz}. The effective temperatures (T$_{\rm eff}$) were
taken from Cohen \& Kuhi (1979) who found this parameter for young
stars with luminosity class V. Surface gravities (log g) for
luminosity class V stars were taken from Strai\v{z}ys \& Kuriliene
(1981). The observed SED (as e.g. discussed in Leinert et al.) rises
from the optical region through the near infrared to 20 $\mu$m. Both
components of the binary are approximately equally bright at 1 $\mu$m
and 1.25 $\mu$m. To match the Kurucz model with the data, the flux of
the G1 Kurucz model atmosphere must be multiplied by 2.25 to get both
stars equally bright at those wavelengths (see figure~\ref{kurucz}).
%This was found to an accuracy of 12 \%. 

By taking these ratios it is interesting to see if the above models
resemble the spectrum of HK Ori. This was achieved by combining the
spectrum of an A dwarf with a G dwarf (multiplied by 2.25) at
wavelengths from 5700 to 6800 $\rm \AA$. The most appropriate stars
for this within the EXPORT data are 24 CVn (A4 V) and 51 Peg (G2.5 V).
These were chosen because they have the same, or very nearly the same,
spectral types as an A4 or G1 star.  Figure~\ref{combo} shows the
spectrum of HK Ori (top) along with a combined spectrum of 24 CVn and
51 Peg (2nd from top), and for comparison the spectrum of 24 CVn (2nd
from bottom) and 51 Peg (bottom). This combined spectrum compares very
well with HK Ori and also reveals how the G star absorption lines are
quite strong in this wavelength region.  The spectral classification
given by Mora et al. (2001) of a G1 type star, in the wavelength
region of 5700 to 6800 $\rm \AA$, has an uncertainty of five spectral
subtypes. Repeating the above but replacing the G1 type star with an
F6 or G6 type star does not significantly change the overall
comparison with the spectrum of HK Ori. 
% This is because a G6 type
%star has very strong absorption lines within this wavelength region,
%however its total flux is less than that of a G1 type star, which when
%combined with an A4 type star in the appropriate ratios, produces a
%very similar comparison spectrum with HK Ori. This also happens for
%the F6 type star which has weaker absorption lines but a stronger
%flux. 

The flux of the G star Kurucz model atmosphere must be increased to
fit the models and observations of HK Ori.  Assuming a constant
temperature, the only parameter that causes an increase of the
luminosity is a larger radius of the star. An increased radius, in this
case 50\%, compared to that of a normal main sequence G star is not
unexpected if the G star is in fact a PMS star  still in the phase
of contraction.

\section{Results - Spectro-astrometry}

\begin{figure}
\mbox{\epsfxsize=0.48\textwidth\epsfbox{ 
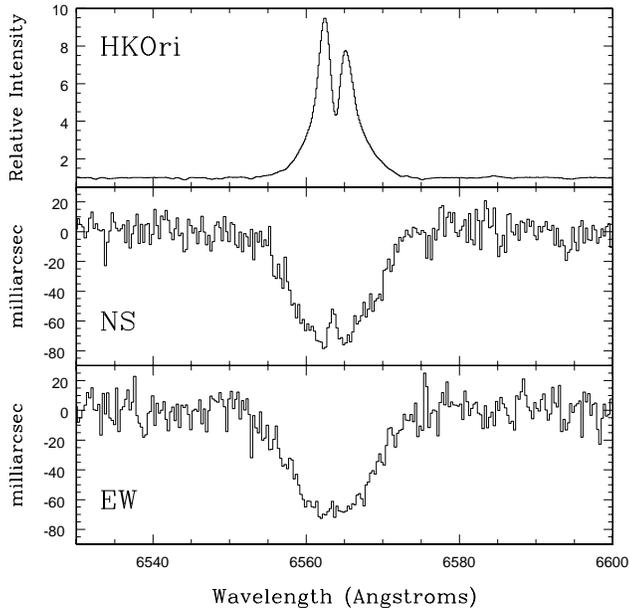}} 
\caption{ Spectro-astrometry of the HK Ori system.  Across the
H$\alpha$ emission of HK Ori, the star responsible for the emission is
detected, while the continuum traces both objects equally well. North and East 
are up in the respective graphs. 
The position angle measured from the EW and NS 
offset spectra corresponds to within a degree with Leinert et al. (1997).  
The rms of the position spectrum is 10 mas. 
\label{specast}
}
\end{figure}

Fig.~\ref{specast} shows the spectro-astrometric data of HK Ori. The
upper panel shows the intensity spectrum centred around H$\alpha$.
Below this are the North-South (NS) and East-West (EW) components of
the position spectrum derived from a set of observations at slit
position angles 0$^{\rm o}$, 90$^{\rm o}$, 180$^{\rm o}$, and
270$^{\rm o}$. One can clearly see a spectro-astrometric signature
occurring across the same wavelength region as the H$\alpha$ emission
line. The two position spectra show a typical binary signature; the
shift in the position spectrum is due to the fact that the H$\alpha$
emitting star is dominant at those wavelengths. This is the first,
independent, confirmation of the binary status of HK Ori.

The feature in the NS position spectrum extends to almost 80 mas and
in the EW to $\sim$70 mas, giving an observed difference of 106
mas. This is a lower limit to the true binary separation since we 
measure the difference between the position of the continuum of one
of the stars and the position of a gaussian fit to the combined
continua of the two stars in the system.  The binary separation of
this system was found to be 340 $\pm$ 20 mas by Leinert et al. (1997),
consistent with the lower limit derived from the position spectra in
Fig.~\ref{specast}.

\begin{figure}
\mbox{\epsfxsize=0.48\textwidth\epsfbox{ 
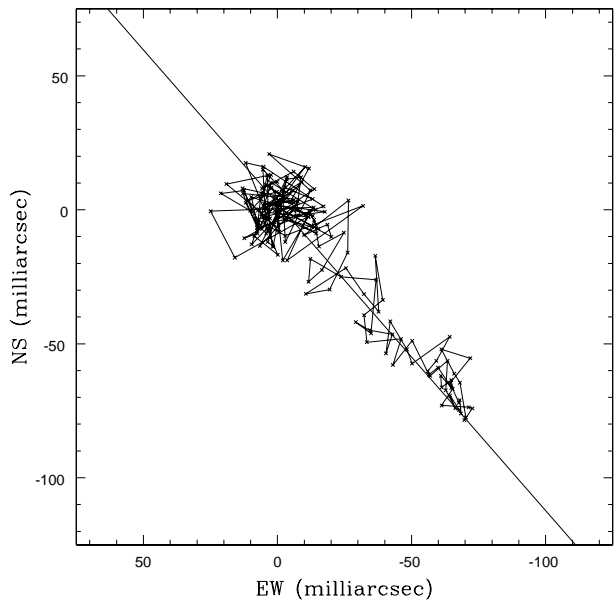}}
\caption{The XY plot of the spectro-astrometry data over the H$\alpha$ profile
for HK Ori. The straight line through the data is the binary position
angle. North is at the top of the plot and East to the left. 
The continuum position is at 0,0. 
\label{specast2d}
}
\end{figure}

In Fig.~\ref{specast2d} the EW position values  are plotted
against the NS position values to obtain a two-dimensional plot
showing the position change through the H$\alpha$ emission line (from
6540 to 6590 {\rm \AA}). Because the excursion in Fig.~\ref{specast2d}
from the continuum (measuring the emission from both stars)
to the H$\alpha$ emission (tracing the line emission) moves in the
South-Western direction, the South-Western star is the H$\alpha$
dominating star.  The direction of the excursion provides the binary
position angle.  From a weighted fit to the points, plotted as a
straight line in the graph, we determine the binary position angle to
be 41.1 $\pm$ 1.0$^{\rm o}$.  This is  close to that of
Leinert et al. (1997) who measured a position angle of 41.7 $\pm$
0.5$^{\rm o}$.

The data do not allow us to straightforwardly determine with certainty
which of the two binary components dominates the optical light
however.  If the flux difference between both stars were large, then
the centroid position of the combined spectrum would be skewed towards
the position of the brightest star. If the optically brightest object
were the H$\alpha$ emitter, the centroid position across H$\alpha$
would then hardly differ from its continuum value. If on the other
hand, the fainter object was emitting in H$\alpha$, the centroid
position will differ significantly from its continuum value and a
large excursion would be detected in the figures.

In the case of HK Ori the situation is not that clearcut
however. Although the A star outshines the G star in the optical
(e.g. Fig.~\ref{kurucz}) it is only 1.5 times brighter than the G star
in the continuum nearby H$\alpha$. The centroid position of the
combined continuum is therefore roughly halfway between both objects,
and the excursion across H$\alpha$ in the position spectrum would be
about equally large irrespective whether the optically brightest or
the faintest object emits in H$\alpha$.  In order to resolve this
issue in the case of HK Ori, dedicated  models are required to
simulate the data. This will be the subject of a forthcoming paper. 

In principle, one could apply the same method to photospheric
absorption lines to determine which star is associated with the
H$\alpha$ emission. However, no obvious signatures across absorption
lines were found in the present data. We attempted co-adding
individual spectral lines to improve the signal-to-noise but subtle,
unrepeatable instrumental effects at milli-arcsec level currently
prevent us from straightforwardly applying this method to weaker and
narrower lines.

In summary, the position plot (Fig.~\ref{specast2d}) indicates that
the South-Western component of the binary is the H$\alpha$ dominant
object, but we are not yet able to determine whether this is the G or
A type star. Leinert et al. (1997) found that this object is also
dominating the {\it H} and {\it K} bands' emission. This excess
emission is very red. At {\it J} both objects are roughly equally
bright, but the {\it J$-$K} colour of the infrared bright object
$\sim$2.1, indicates a large, red excess.  Now, a clearer picture is
starting to emerge. The binary component, that emits most of the
H$\alpha$, also dominates the near-infrared emission and is clearly
the (most) active, young star.  The other component is less active, it
does not radiate much H$\alpha$ emission, and is faint in the
near-infrared. In the following we will argue that the system is
composed of an active late type T Tauri star with a young, less active A
star, which is probably settled, or is settling, on the main sequence.

\section{Discussion and concluding remarks}

\subsection{The pre-main sequence nature of the system}

The spectrum of HK Ori reveals a complex behaviour, ranging from
double peaked hydrogen recombination lines to variable inverse P Cygni
profiles in the helium lines.  The different lines probe different
regions surrounding the star. The helium emission and absorption
predominantly originates in the hot shocked region where material has
accreted onto the star (e.g. Beristain, Edwards \& Kwan 2001), whereas
the redshifted Na transient absorption comes from the neutral material
further from the star (see e.g. de Winter et al. (1999) on BF Ori).
The double peaked emission line profiles of both the H$\alpha$ and
H$\beta$ lines may imply formation of the lines in a Keplerian
circumstellar disc as the peak separation of the H$\beta$ emission is
larger than that of H$\alpha$. In principle, spectro-astrometry should
be able to detect the presence of a rotating disk, as the blue- and
redshifted emission peaks would be located at either side of the
stellar continuum. It turns out that we need a much better spatial
resolution than used here to be able to resolve such disks (Baines
2004).  Having said that, rotation does not explain the shift of about
20 kms$^{-1}$ that is observed in the January data. This may well be
the result of the increased activity of the star during that period,
as evidenced by the extra veiling for example.  Indeed, to all intents
and purposes, the spectrum resembles that of a low mass classical T
Tauri star undergoing variable accretion (see for example the review
by Bertout, 1989).  The line variability and variable veiling can be
explained by magnetospheric accretion. In this model the stellar
magnetic field disrupts the circumstellar accretion disc and accreting
material proceeds to the stellar surface by falling along the magnetic
field lines (see e.g. Wood et al., 1996; Muzerolle, Hartmann \&
Calvet, 1998). The rotating accretion disk may not be observable in
high excitation lines such as helium, as the truncation of the inner
disk may prevent this highly excited material being present in a
rotating geometry.

Although the A star dominates the optical continuum, evidence that the
spectral properties are due to the active T Tauri star is also provided
by the fact that the system suffers from variable veiling - the
spectrum is brighter and bluer in January 1999, and shows the same
absorption and emission lines, only slightly weaker - again a well
known observational property of T Tauri stars. The presence of the Li
{\sc i} 6708 $\rm \AA$ line is a characteristic of young low mass
stars (Corporon \& Lagrange, 1999). An additional observation is that
provided by Goodrich (1993) who found a Herbig-Haro object associated
with HK Ori, objects which are predominantly associated with lower
mass stars.

In addition to the usual T Tauri characteristics, other properties are
common to both Herbig Ae and T Tauri stars. As both the photometric and
polarimetric variability appear to be correlated, it is likely that
this pre-main sequence star (or at least one of the stars in the
binary system) is an UXOR (first suggested by Oudmaijer et al.,
2001). The star appears to be viewed approximately edge-on through a
clumpy, dusty circumstellar disc. Occasionally large dust clouds,
orbiting within the circumstellar disc, obscure the star and its
H$\alpha$ emitting envelope.

\subsection{The binary}

The spectroscopic data confirm previous suggestions that HK Ori is a
binary system consisting of a hot and a cool component. Specifically,
the data appear to be very well represented by a combination of an A
dwarf star spectrum superposed on a (sub-)giant G-type spectrum.  The
radial velocity of the system is consistent with previous
determinations, and does not reveal binary motion, as expected for a
wide binary. The spectro-astrometry of HK Ori confirms, for the first
time, the speckle data from Leinert et al. (1997).  The position angle
was derived to within a degree of that measured from the speckle data,
while a lower limit to the true separation is measured by the shift in
the photocentre between line and continuum. However, the true
separation can be recovered with dedicated simulations.  We note that
the binary detection was by no means trivial: Pirzkal, Spillar \& Dyck
(1997) published a paper with similar near-infrared speckle
observations as those by Leinert et al., but were unable to detect a
companion to HK Ori.

The spectro-astrometric data indicate that the infrared bright
component of the binary is also the H$\alpha$ emitter. Based on the
fact that the spectral properties resemble that of a T Tauri star, we
suggest that this component is the G star, and that therefore 
this object is the optically fainter one.

\subsection{Final remarks}

In this paper we have investigated the pre-main sequence binary HK
Ori. The combination of multi-epoch data and broad wavelength coverage
spectroscopy as well as the dedicated spectro-astrometric observations
have provided very useful information on both components in this
otherwise unresolved system and are complementary to existing speckle
data.

There is a realistic hope of learning significantly more about such
systems from spectro-astrometry when model simulations, optimized to
investigate longslit spectra of sub-arcsecond binaries, become
available.  Improved data will be crucial for our understanding of
pre-main sequence evolution, especially now that binary systems seem
to be the norm rather than the exception in young stars (e.g. Leinert
et al. 1997, Prato \& Simon 1997).  The natural next step will be to
obtain system parameters such as the separation and the flux difference
for both stars from the spectro-astrometry alone. The ultimate goal is
to be able to properly distinguish the spectra from the respective
stars. and derive physical parameters of both components.  These
issues will be discussed in a future paper (Porter, Oudmaijer \&
Baines 2004).

The current generation of telescopes should allow binary systems with
much smaller separations than that of HK Ori (0.34 arcsecond) to be
studied, certainly if equipped with adaptive optics.  We have
illustrated in this paper that spectro-astrometry can routinely detect
subarcsecond binaries even in comparatively bad seeing of 2 arcsec,
whilst not requiring any special techniques. Based on the relative
faintness of HK Ori ({\it V}$\sim$11), observations of brighter
objects or at higher signal-to-noise should enable the detection of
binary companions at milli-arcsecond scales.

\vspace*{-0.5cm}

\section*{Acknowledgements}

Debbie Baines acknowledges support from a PPARC student grant. The
work of C. Eiroa, B. Mer\'{\i}n, B. Montesinos, A. Mora and E. Solano
has been supported in part by the Spanish grant AYA2001-1124-C02.

\label{lastpage}


\vspace*{-0.5cm}

\begin{thebibliography}{}
\bibitem{b1} Bailey J., 1998, MNRAS, 301, 161
\bibitem{b1} Baines D., 2004, PhD thesis, University of  Leeds, U.K. 
\bibitem{b1} Beristain G., Edwards S., Kwan J. 2001, ApJ 551, 1037
\bibitem{b3} Bertout C., 1989, ARA\&A, 27, 351
\bibitem{b4} Bouvier J., Corporon P. 2001, IAU Symposium 200, ASP, ed. Zinnecker H, Mathieu R.D. p. 155  
\bibitem{b5} Cohen M.,  Kuhi L. V., 1979, ApJS, 41, 743 
\bibitem{b9} Corporon P., Lagrange A. M., 1999, A\&AS, 136, 429  
\bibitem{b11} de Winter, D., Grady, C. A., van den Ancker, M. E., et al, 1999, A\&A, 343, 137  
\bibitem{b15} Eiroa C. (EXPORT) 2000, in Disks, Planetesimals and Planets, ed. F. Garz\'on, C. Eiroa, D. de Winter \& T. Mahoney, ASP Conf. Ser. 219, 3
\bibitem{b15} Eiroa C., Garz\'on F., Alberdi A., et al, 2001, A\&A 365, 110
\bibitem{b15} Eiroa C., Oudmaijer R.D., Davies J.K., et al, 2002, A\&A 384, 1038
\bibitem{b15} Finkenzeller U., Jankovics I., 1984, A\&AS, 57, 285
\bibitem{b16} Finkenzeller U., Mundt R., 1984, A\&AS, 55, 109
\bibitem{b17} Goodrich R.W., 1993, ApJS, 86, 499
\bibitem{b18} Grady C.A., P\'{e}rez M.R. Talavera A., et al., 1996, A\&AS, 120, 157  
\bibitem{b19} Grinin V.P., Kolotilov E.A., Rostopchina A., 1995, A\&AS, 112, 457
\bibitem{b22} Herbig G.H., 1960, ApJS, 4, 337
\bibitem{b23} Herbst W., Shevchenko V.S., 1999, AJ, 118, 1043 
\bibitem{b25} Joy A.H., 1949, ApJ, 110, 424  
\bibitem{b26} Kurucz, R. L., 1993, CD-ROM 1-23 Smithsonian Astrophysical Observatory
\bibitem{b27} Leinert C., Richichi A.,  Haas M., 1997, A\&A, 318, 472
\bibitem{b28} Mora A., Mer\'in B., Solano E., et al, 2001, A\&A, 378, 116
\bibitem{b29} Muzerolle J., Hartmann L., Calvet N., 1998, AJ, 116, 455
\bibitem{b31} Oudmaijer R. D., Palacios, J., Eiroa, C., et al, 2001, A\&A 379, 564 
\bibitem{b32} Pirzkal N., Spillar E.J., Dyck H.M., 1997, ApJ, 481, 392 
\bibitem{b31} Porter J., Oudmaijer R.D., Baines D. 2004, A\&A, submitted
\bibitem{b32} Prato L., Simon M., 1997, ApJ, 474, 455
\bibitem{b32} Reipurth B., Pedrosa A., Lago M.T.V.T., 1996, A\&AS, 120, 229  
\bibitem{b33} Strai\v{z}ys V., Kuriliene G., 1981, Ap\&SS, 80, 353
\bibitem{b38} Takami M.,  Bailey, J. Chrysostomou, A. 2003, A\&A 397, 675
\bibitem{b38} Tohline J.E. 2002, ARA\&A 40, 349
\bibitem{b38} Whelan E.T., Ray T.P., Davis C.J. 2004, A\&A 417, 247
\bibitem{b38} Wood, K., Kenyon, S. J., Whitney, B. A., et al, 1996, ApJ, 458, L79
\end{thebibliography}
\end{document}